\documentclass[intlimits,twoside,a4paper]{article}

\usepackage{amsmath,amssymb}
\usepackage{graphicx}

\usepackage[T2A]{fontenc}
\usepackage[cp1251]{inputenc}

\usepackage{ifthen}
\usepackage{booktabs}
\usepackage{color}

\usepackage{dcolumn}
\usepackage{longtable}
\usepackage{gensymb}

\usepackage[eqsecnum]{cmpj2}

\newcommand{\bb}{\begin{equation}}
\newcommand{\ee}{\end{equation}}
\newcommand{\ba}{\begin{eqnarray*}}
\newcommand{\ea}{\end{eqnarray*}}
\newcommand{\rhor}{\rho(\textbf{r})}
\newcommand{\dd}{{\rm d}}
\newcommand{\rr}{{\mathbf r}}
\newcommand{\dr}{{\rm d}\textbf{r}}

\issue{2016}{19}{1}{13604}
\doinumber{10.5488/CMP.19.13604}

\title[Phase transitions of fluids in heterogeneous pores]%
{Phase transitions of fluids in heterogeneous pores}
\author[A. Malijevsk\'y]{A. Malijevsk\'y\refaddr{label1,label2}}
\addresses{
\addr{label1} Department of Physical Chemistry, University of Chemistry and Technology Prague, \\ 166 28 Praha 6, Czech Republic
\addr{label2} Laboratory of Aerosols Chemistry and Physics, Institute of Chemical Process Fundamentals, \\ Academy of Sciences, 16502 Prague 6, Czech Republic
}

\date{Received November 13, 2015, in final form December 22, 2015}

\begin{document}

\maketitle

\begin{abstract}
We study phase behaviour of a model fluid confined between two unlike parallel walls in the presence of long range (dispersion) forces. Predictions obtained from
macroscopic (geometric) and mesoscopic arguments are compared with numerical solutions of a non-local density functional theory. Two capillary models are considered. For
a capillary comprising of two (differently) adsorbing walls we show that simple geometric arguments lead to the generalized Kelvin equation locating 
capillary condensation very accurately, provided both walls are only partially wet. If at least one of the walls is in complete wetting regime, the Kelvin equation should be modified by
capturing the effect of thick wetting films by including Derjaguin's correction. Within the second model, we consider a capillary formed of two competing walls, so that
one tends to be wet and the other dry. In this case, an interface localized-delocalized transition occurs at bulk two-phase coexistence and a temperature $T^*(L)$
depending on the pore width $L$. A mean-field analysis shows that for walls exhibiting first-order wetting transition at a temperature $T_\text{w}$, $T_\text{s}>T^*(L)>T_\text{w}$, where the
spinodal temperature $T_\text{s}$ can be associated with the prewetting critical point, which also determines a critical pore width below which the interface
localized-delocalized transition does not occur. If the walls exhibit critical wetting, the transition is shifted below $T_\text{w}$ and  for a model with the binding potential
$W(\ell)=A(T)\ell^{-2}+B(T)\ell^{-3}+\cdots$, where $\ell$ is the location of the liquid-gas interface, the transition can be characterized by a dimensionless parameter
$\kappa=B/(AL)$, so that the fluid configuration with delocalized interface is stable in the interval between $\kappa=-2/3$ and $\kappa\approx-0.23$.
\pacs 68.08.Bc, 05.70.Np, 05.70.Fh
\end{abstract}

\section{Introduction}

It is very well known that structure and phase behaviour of a confined fluid is quite distinct from that of its bulk counterpart. A familiar example of a confining
geometry is a slit pore formed by two parallel, identical and infinite plates, a distance $L$ apart. The combination of finite-size effects and fluid adsorption at
the walls leads to a shift in the liquid-vapour phase boundary and in the critical point compared to a bulk fluid \cite{fisher1, fisher2, ev_tar, evans90, gelb}. The
location of this capillary condensation (or evaporation) transition is macroscopically described by Kelvin's equation which predicts that the chemical potential at
which the vapour in the slit condenses into the liquid-like phase is shifted from its saturation value $\mu_\text{sat}$ by an amount
 \bb
 \delta\mu\equiv\mu_\text{sat}-\mu_\text{cc}=\frac{2\gamma\cos\theta}{(\rho_\text{l}
 -\rho_\text{v})L}\,, \label{kelvin}
 \ee
where $\rho_\text{l}$ and $\rho_\text{g}$ are the coexisting bulk liquid and gas densities, respectively, $\gamma$ is the liquid-gas surface tension and $\theta$ is the contact
angle of a macrosocpic liquid droplet sitting on isolated wall. The role of wetting layers adsorbed at the walls when $\theta=0$, i.e., for $T>T_\text{w}$, where $T_\text{w}$ is the
wetting temperature corresponding to a semi-infinite system $L\to\infty$,  has also been appreciated but this effect is mostly quantitative, such that the presence of
the layers effectively reduces the pore width \cite{ev_marc}. If the isolated walls exhibit first-order wetting transition at $T=T_\text{w}$ (and $\mu=\mu_\text{sat}$), a
prewetting transition corresponding to a finite jump in the wetting layers thickness can also occur, although the transition is typically metastable with respect to the
capillary condensation unless the pore width is fairly large \cite{ev_pre}.

The scenario of a liquid-vapour coexistence can, however, be very different in pores made of unlike walls. This was demonstrated by Parry and Evans \cite{par_ev1,
par_ev2} who proposed a model, treated within the Landau theory in the language of magnets, with perfectly antisymmetric surface fields of the walls. They have shown
that, for sufficiently large pores, the capillary condensation transition is replaced by the interface localization-delocalization phase transition which occurs at a
two-phase bulk coexistence at a temperature $T^*(L)$ near the wetting temperature $T_\text{w}$ of the wall with the affinity to ``$+$'' phase (which, owing to the symmetry of
the system, is identical to the wetting temperature of the opposing wall which tends to be wetted by ``$-$'' phase). The transition separates a regime, present for
$T<T^*(L)$, when the equilibrium density profile corresponds to a very thick ``$+$'' or to a very thick ``$-$'' phase with a ``$+-$'' interface pinned to either of the
walls, from the high temperature regime at which the interface is unbounded from either of the walls. In the latter case, sometimes referred to as soft-mode phase, the
interface finds a compromise between the  antagonistic wetting preferences of the walls, such that it develops around the midpoint of the pore being a subject of large
fluctuations causing the interface to wander along the pore. Compared with the former model of pores with identical walls, the case of antisymmetric walls makes much
closer link with wetting properties of the walls. In the theory of wetting phenomena, the concept of a binding potential proved to be very useful (at least on a
mean-field level) and can be used to describe the phase behaviour of the fluid confined between antisymmetric walls. If the walls, when isolated, exhibit critical
wetting as assumed in references \cite{par_ev1, par_ev2}, the binding potential at each wall acquires a single minimum whose location shifts continuously from the wall to
infinity as the temperature increases towards $T_\text{w}$ along the phase coexistence line [$\mu=\mu_\text{sat}(T)$]. It means that, when the pore is sufficiently wide, the
binding potential has two minima that are of the same depth and are located symmetrically  around the midpoint of the pore. There are thus two equally stable solutions
for the density profile corresponding to large and small adsorptions. The adsorption difference, or equivalently the distance between the two minima, decreases with
an increasing temperature and ultimately disappears at $T^*(L)$, which is a finite-size shift of $T_\text{w}$. Above $T^*(L)$, the binding potential landscape adopts a $U$-shape
with a very shallow single minimum which enables the interface to drift around the centre with a very small free energy cost.

In contrast to pores with identical walls, the nature of wetting transition at the (isolated)  walls becomes much more important when the walls are antisymmetric. If
they exhibit first-order wetting, the characteristic feature of the binding potential is a competition between a minimum at a finite distance from the wall with the
unbounded state corresponding to a minimum at infinity. Assuming the pore is sufficiently wide, the binding potential now possesses three local minima, such that two
minima near the walls are the global minima for temperatures below the finite-size shifted wetting temperature $T^*(L)$, whilst the middle minimum becomes a global
minimum for $T>T^*(L)$. The nature of the transition (at fixed $L$ and $T$ varying) thus reflects the nature of wetting at the walls: while the transition is continuous
for a critical wetting, in which case two potential minima continuously merge in the middle of the pore, it becomes discontinuous for first-order wetting due to a jump
in the location of the global minimum. Consequently, $T^*(L)$ is a critical point for critical wetting, whereas it is a triple point for first-order wetting. However,
all these conclusions are only valid for sufficiently large pores. By decreasing the pore width, the space to accommodate all three minima of the binding potential  is
reduced and when the middle minimum disappears, the order of the transition becomes second order \cite{indekeu, binder4}.

These predictions have been verified by extensive Monte Carlo simulations by Binder {et al.} for Ising-like models \cite{binder1, binder2, binder3, binder_rev,
binder5, binder_asym}. More recently, the properties of the interface localization-delocalization transition were also studied for fluid models of soft matter systems
\cite{binder_soft1, binder_soft2, binder_soft3}. Compared to magnets, however, the situation with fluids is a bit more intricate. Firstly, owing to unequal entropy of
coexisting phases, they lack the perfect symmetry of Ising-like models which makes the concept of antisymmetric walls less clear. Secondly, ubiquity of dispersion forces
in fluid systems is an important extra ingredient to be considered which, in fact, prevents complete drying. The latter problem can be avoided by considering binary
(colloid-polymer)  mixtures such as in references~\cite{binder_soft1, binder_soft2, binder_soft3}. The long-range dispersion interaction was included in reference \cite{stewart}
by Stewart and Evans for a simple, one-component fluid; in this study, extensive density functional theory (DFT) calculations have been made to confirm scaling
predictions for several thermodynamic quantities. The wall parameters have been set such that they ensure a complete wetting on one wall and a complete drying on the
opposite wall and that, for the given fixed subcritical temperature, the corresponding Hamaker constants, i.e., the coefficients of the lowest order term in the
respective binding potentials, are identical. However, even if neglecting the higher order contributions in the binding potential, the system does not exhibit the same
``antisymmetry'' as in the case of magnets. This is because the potential exerted by both walls must contain a repulsive part for the walls to be impenetrable, but
only one of them (the ``solvophilic'' wall) also contains an attractive portion.  This produces a binding potential well so that there exists a wetting temperature $T_\text{w}$
below which the wall is only partially wet. By contrast, the other, ``solvophobic'' wall is purely repulsive and thus dried (i.e., wet by gas) at all temperatures.
Consequently, the binding potential landscape for a fluid confined between these competing walls is clearly different from that described above for magnetic systems.
Furthermore, since the Hamaker constants depend on a temperature (via coexisting densities), any change in the temperature would necessarily break the ``antisymmetry''
of the system, which is thus not a property of the system itself but also depends on thermodynamic parameters.

In this paper, we study the phase behaviour of a fluid confined between asymmetric walls that  interact with the fluid via dispersion forces. There are two models that we
use to represent such a heterogeneous pore. Within the first model, both walls exhibit the wetting transition but at different temperatures. We present simple geometric
arguments that lead to the extended Kelvin equation which predicts the location of capillary condensation in the pore. For the case, when the temperature of the system
is greater than the wetting temperature of at least one of the walls, we modify Kelvin's equation by incorporating the effect of the presence of the wetting layer(s).
Both of these predictions are tested against numerical results obtained from a non-local density functional theory (DFT). The second model represents the case of
antisymmetric walls, so that one of the walls is wetted by liquid whereas the other wall by gas. We present a mean-field analysis for the location of the interface
localization-delocalization transition. The comparison with DFT shows that the analytic predictions are surprisingly accurate down to very narrow pores at least for the
case, when one wall exhibits first-order wetting transition (as opposed to critical wetting) and the other wall is completely dried (wetted by gas) at all temperatures.

The remainder of the paper is organized as follows. In section~\ref{sec:2} we present geometric  arguments to determine the location of capillary condensation for pores with two
unlike walls and also show how this extended Kelvin's equation can be modified to embrace Derjaguin's correction due to the presence of wetting layers. In this section
we also consider the pore model consisting of antisymmetric walls for which we present a mean-field analysis to locate the interface localization-delocalization
transition. In section~\ref{sec:3} we set the molecular model and show the DFT results. Although the planar symmetry permits to treat the system as a one-dimensional problem, we
employ a two-dimensional DFT to test the plausibility of geometric arguments. We then examine the properties of the interface localization-delocalization transition for
antisymmetric walls by determining the binding potential and study its behaviour as the temperature and pore width vary. We summarise and discuss our results in the
concluding section~\ref{sec:4}.

\section{Heuristic arguments}

\label{sec:2}

\begin{figure}[!t]
\centerline{
\includegraphics[width=2.5cm]{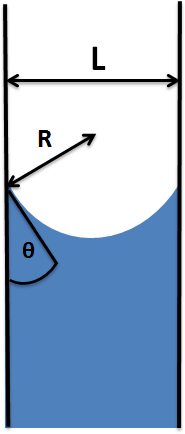}
\hspace*{3cm}
\includegraphics[width=3cm]{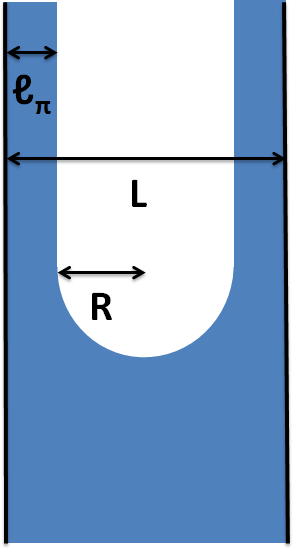}
}
\caption{(Color online) A schematic picture illustrating a coexistence of liquid-like and  gas-like phases in a homogeneous pore of a macroscopic width $L$. In the left panel, a
contact angle at the walls is assumed to be $\theta>0$ meaning that the temperature of the system is below the wetting temperature. The radius $R$ of the cylindrical
meniscus separating the phases is given by the Laplace pressure $R=\gamma/\delta\mu(\rho_\text{l}-\rho_\text{v})$. In the left panel, the contact angle is zero, so that the meniscus
meets tangentially the wetting layers of thickness~$\ell_\pi$.} \label{slit}
\end{figure}

We start by recalling macroscopic arguments leading to a condition for a liquid-vapour equilibrium in a homogeneous pore, i.e., a pore made of identical walls that
exhibit a wetting transition (by liquid) at a temperature $T_\text{w}$. Within the purely macroscopic treatment, the distance between the walls $L$ is taken to be large and it is
assumed that a first-order transition between a state corresponding to a gas-like and a liquid-like phase occurs for any temperature $T$ below the bulk critical
temperature $T_\text{c}$. We then expect that the pressure $p$ (or the chemical potential $\mu$) at which the transition occurs in the pore is shifted below the saturation
value $p_\text{sat}$ (or $\mu_\text{sat}$). Based on the surface thermodynamics, this value can be determined by a simple free energy balance for a low- and a high-density
state which leads to the well known Kelvin's equation (\ref{kelvin}). Apart from this thermodynamic picture, Kelvin's equation has also a geometric interpretation as it is
illustrated in figure~\ref{slit}. Based on this approach, one considers a single pore in which both phases coexist. Since the equilibrium occurs off bulk two-phase
coexistence, the interface between the gas and vapour phases  in the pore must be curved in the direction perpendicular to the walls with a Laplace radius
$R=\gamma/\delta p$, where $\delta p=p_\text{sat}-p$. Assuming that $\delta p$ is small, we can write $\delta p=\delta\mu(\rho_\text{l}-\rho_\text{v})$ according to the Gibbs-Duhem
relation, where $\rho_\text{l}$ and $\rho_\text{v}$ are the particle densities of coexisting liquid and vapour phases, respectively. Substituting from $R=L/(2\cos\theta)$ yields
equation~(\ref{kelvin}).

Kelvin's equation has proven to be fairly accurate down to surprisingly small values of  $L$ for temperatures $T<T_\text{w}$. Above the wetting temperature, when liquid layers
adsorb at the walls, the appropriate geometric picture of the phase coexistence in the pore is shown in the right-hand panel of figure~\ref{slit}. In this case, the
cylindrically-shaped meniscus tangentially connects the wetting films that are of thickness $\ell_\pi$. For sufficiently wide pores, $\ell_\pi$ can be considered as a
thickness of a complete wetting layer adsorbed on an isolated wall. This geometric interpretation  suggests that the pore width $L$ appearing in the denominator of
equation~(\ref{kelvin}) should be replaced by its effective value $L_\text{eff}=L-2\ell_\pi$. Explicit calculations based on a free energy balance including the effective
interaction between interfaces show that $L_\text{eff}$ depends on the range of the molecular forces and that $L_\text{eff}=L-3\ell_\pi$ when long range nonretarded
dispersion forces are involved.

\begin{figure}[!t]
\centerline{
\includegraphics[width=2.5cm]{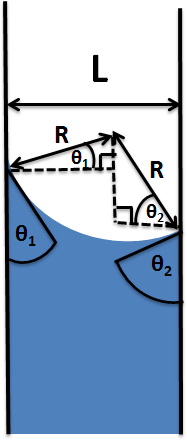}
\hspace*{3cm}
\includegraphics[width=3cm]{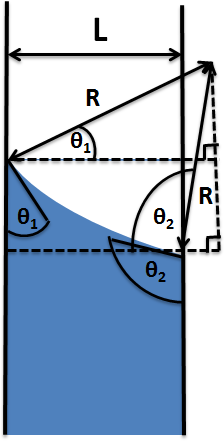}
}
\caption{(Color obline) Sketch of a vapour-liquid coexistence in a heterogeneous pore of a width $L$. In this picture, the left-hand wall is partially wet and the right-hand wall is either partially wet but with a larger contact angle (left-hand panel) or
partially dried ($\theta_2>\pi/2$, as shown in the right-hand panel). Accordingly, the centre of the meniscus of a radius $R=\gamma/\delta\mu(\rho_\text{l}-\rho_\text{v})$ is now closer to the right-hand wall.} \label{slit_het}
\end{figure}

We now turn our attention to a heterogeneous pore, where, within the macroscopic treatment, the walls are characterised by contact angles $\theta_1$ and $\theta_2$. For
the rest of the paper we shall assume without any loss of generality that the contact angle at the left-hand wall $\theta_1$ is not larger than $\theta_2$ and that
$\theta_1\leqslant \pi/2$ [if $\theta_1>\pi/2$, the problem is either reversed spatially (if $\theta_2<\pi/2$) or thermodynamically (if $\theta_2>\pi/2$) in which case the gas
and liquid phases interchange their roles]. We can again construct a geometric picture of the liquid-vapour coexistence in the pore, as is shown in figure~\ref{slit_het}.
Now, the center of the meniscus is no more in the centre of the pore but is shifted towards the right-hand wall or even beyond in case of $\theta_2>\pi/2$.  A simple geometry
then leads to the following generalization of Kelvin's equation for heterogeneous pores~\cite{par_ev1}:
 \bb
 \delta\mu\equiv\mu_\text{sat}-\mu_\text{cc}=\frac{\gamma(\cos\theta_1+\cos\theta_2)}{(\rho_\text{l}-\rho_\text{v})L}\,. \label{kelvin_het}
 \ee

Usually, Kelvin's equation is interpreted such that it tells us what is the chemical potential at the capillary condensation for a given pore width $L$. This view can also be reversed and we can ask what is the equilibrium
distance $L$ when we fix $\delta\mu$ and $T$. From the geometric construction shown in figure~\ref{slit_het} it is clear that the ``more heterogenous'' the pore is, the shorter the distance $L$ must be for thermodynamic
criteria to be met.

Next, we wish to adopt a more microscopic view for the case when either $\theta_1=0$ or $\theta_2=0$ (or both). We then expect that the wetting layer of a thickness
$\ell_\pi^{(i)}$ which is formed at the wall $i$ will somewhat modify the purely macroscopic prediction of equation~(\ref{kelvin_het}). When a wetting layer of thickness
$\ell_\pi^{(i)}$ intrudes between a single wall $i$ and the bulk gas, the corresponding surface free energy of the wall-gas interface is as follows:
 \bb
 \gamma_{\text{w}_{i}\text{g}}=\gamma_{\text{w}_{i}\text{l}}+\gamma+W_i\left(\ell_\pi^{(i)}\right),\label{swg}
 \ee
where $\gamma_{\text{w}_{i}\text{l}}$ is the surface tension between the wall $i$ and the liquid, and
  \bb
  W_i(\ell)=\delta p\ell+\frac{A_{i}}{\ell^2}+\frac{B_{i}}{\ell^3}+\cdots\label{W}
  \ee
is the effective potential between the wall-liquid and liquid-gas interfaces due to the wetting layer of a thickness $\ell$. The coefficient $A_i$ is called the Hamaker
constant which must be positive for $T>T_\text{w}$. This form of the binding potential assumes that the wall-fluid or fluid-fluid interactions are dominated by nonretarded
dispersion forces at large distances. The minimum of $W$ determines the equilibrium thickness of the wetting layer
  \bb
  \ell^{(i)}_\pi\approx\left(\frac{2A_{i}}{\delta p}\right)^{1/3}, \qquad T\geqslant T_w^{(i)}\,.\label{ell}
  \ee

  If only a microscopic wetting layer forms at the wall, i.e., if $T<T_w^{(i)}$, then the global minimum of $W_{i}(\ell)$ is negative and the liquid-vapour interface is
  pinned to the wall to a microscopic distance $\ell^{(i)}_\pi$ even for $\delta p=0$, in which case the comparison of (\ref{swg}) with Young's equation reveals that
  \bb
  W_i(\ell_\pi^{(i)})=\gamma(\cos\theta_i-1)\,,\qquad T<T_w^{(i)}\,.\label{bound}
  \ee

Now, in the pore of a large width $L$, we assume that the thickness of the wetting layer adsorbed at either wall  is the same as the one on a single wall. The
free-energy difference per unit area between a low-density state (with the grand-potential per unit area $\omega_\text{g}=-pL+\gamma_{\text{w}_1\text{g}}+\gamma_{\text{w}_2\text{g}}$) and a high-density
state (with the grand-potential per unit area $\omega_\text{g}=-p_\text{l}^+L+\gamma_{\text{w}_1\text{l}}+\gamma_{\text{w}_2\text{l}}$, where $p_\text{l}^+$ denotes a pressure of the metastable liquid at a given
$\mu<\mu_\text{sat}$ and $T$) is then given by
  \bb
  \Delta\omega=\omega_\text{g}-\omega_\text{l}=(p_\text{l}^+-p)L+2\gamma+W_1(\ell_\pi^{(1)})+W_2(\ell_\pi^{(2)})\,,\label{domega}
  \ee
where we have neglected effective interactions other than between the liquid-gas interface and the nearest wall.

In particular, when $\theta_1=0$ and $\theta_2>0$, a substitution of equations (\ref{W}) and (\ref{bound}) into (\ref{domega}) gives
 \bb
 \Delta\omega=(p_\text{l}^+-p)L+\delta p\ell_\pi^{(1)}+\frac{A_{1}(T)}{\left(\ell_\pi^{(1)}\right)^2}+\gamma(1+\cos\theta_2)\,.
  \ee
  Identifying $\delta p\approx p-p_\text{l}^+\approx\delta\mu(\rho_\text{l}-\rho_\text{g})$ to first order in $\delta\mu$ and using equation~(\ref{ell}), we obtain
  \bb
  \delta\mu=\frac{\gamma(1+\cos\theta_2)}{\left(L-\frac{3}{2}\ell_\pi^{(1)}\right)(\rho_\text{l}-\rho_\text{g})}\,.
  \ee

This result can be generalised as follows:
  \bb
  \delta\mu=\frac{\gamma(\cos\theta_1+\cos\theta_2)}
  {\left[L-\frac{3}{2}\left(\ell_\pi^{(1)}+\ell_\pi^{(2)}\right)\right]
  (\rho_\text{l}-\rho_\text{g})}\,,\label{kelvin_het2}
  \ee
 where $\ell_\pi^{(i)}=0$ if $\cos\theta_i>0$.

The case of ``antisymmetric'' walls when $\cos\theta_1=-\cos\theta_2$ deserves special attention.  The modified Kelvin equation implies that the only phase transition
can occur at $\mu=\mu_\text{sat}$. In this case, apart from the gas-like and liquid-like states, a configuration consisting of a thick film of liquid of width $\ell$ and
a film of gas of thickness $L-\ell$ with the liquid-gas delocalized interface parallel to the walls should be also taken into account. The appropriate grand potential
per unit area of this configuration is:
 \bb
 \omega_\text{deloc}(\ell)=\gamma_{\text{w}_1\text{l}}+\gamma
 +\gamma_{\text{w}_2\text{g}}+\frac{A_1}{\ell^2}+\frac{A_2}{(L-\ell)^2}+\frac{B_1}{\ell^3}+\frac{B_2}{(L-\ell)^3}\,,\label{om_deloc1}
 \ee
 where the higher-order terms were neglected.

\begin{figure}[!b]
\centerline{
\includegraphics[width=10cm]{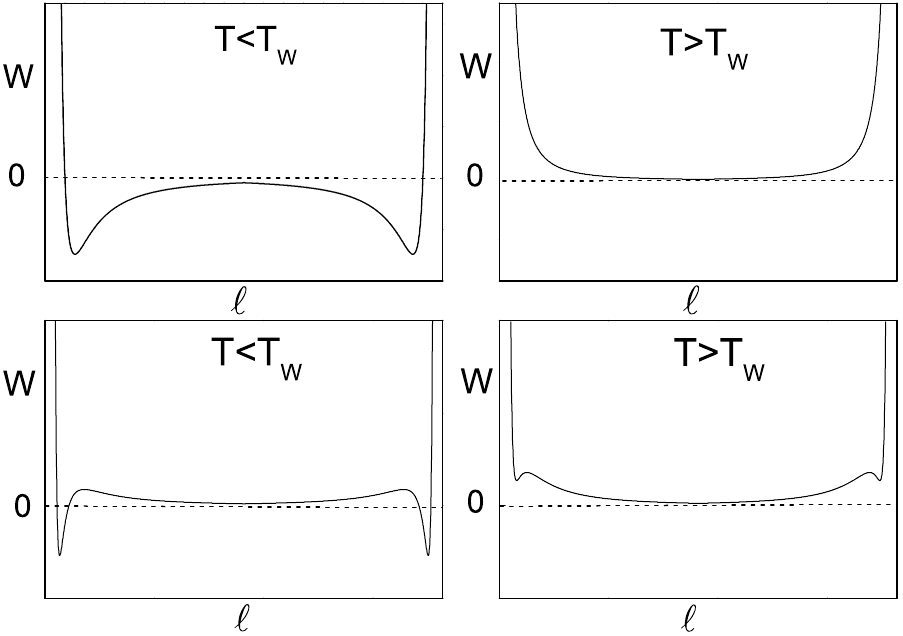}
}
\caption{Schematic behaviour of a binding potential in a macroscopically wide pore with perfectly antisymmetric walls. Below the wetting temperature $T_\text{w}$, the binding
potential has two minima located near the walls that correspond to a high-density and a low-density state in which case the pore is predominantly filled by one phase
with only a microscopic layer of the other fluid phase adsorbed on one of the walls. For $T>T_\text{w}$, there is a single minimum of the biding potential in the middle of the
pore, so that thick liquid-like and gas-like films form at respective walls. The upper panel describes the case when the
walls exhibit critical wetting and the lower panel is appropriate for first-order wetting.} \label{bp_sketch}
\end{figure}

Assuming that the walls are perfectly opposite, $A_1=A_2\equiv A$ and $B_1=B_2\equiv B$ we separately consider the cases $T<T_{\text{w}_1}=T_{\text{w}_2}\equiv T_\text{w}$ and $T>T_\text{w}$. The
respective binding potentials are shown in figure~\ref{bp_sketch} for the walls that undergo critical (upper panels) and first-order (lower panels) wetting transitions, when
$L$ is macroscopic. From the inspection of figure~\ref{bp_sketch} we can conclude that for $L$ macroscopic $T_\text{w}$ represents a triple point for first-order wetting in which
case three minima of $W(\ell)$ corresponding to low-density, high-density and delocalized states are of equal depths. For critical wetting, there are only two minima in
the binding potential below $T_\text{w}$. Upon increasing the temperature, the two minima are getting closer to each other continuously and finally merge at the midpoint of the
pore at $T_\text{w}$ which thus represents a critical temperature in this case.

We now wish to include the finite-size effects due to finite values of $L$, still considering the binding potential of the form (\ref{W}). The grand potential of the delocalized state with the liquid-gas interface in the
middle of the pore is as follows:
 \bb
 \omega_\text{deloc}=\gamma_{\text{w}_1\text{l}}+\gamma+\gamma_{\text{w}_2\text{g}}+\frac{8A}{L^2}+\frac{16B}{L^3}\,,\label{om_deloc}
 \ee
 whereas the grand potentials of the low-density configuration
 \bb
 \omega_{\rm g}=\gamma_{\text{w}_1\text{g}}+\gamma_{\text{w}_2\text{g}} \label{om_low}
 \ee
 and the high-density configuration
 \bb
 \omega_{\rm l}=\gamma_{\text{w}_1\text{l}}+\gamma_{\text{w}_2\text{l}} \label{om_high}
 \ee
are identical by symmetry. For $T<T_\text{w}$, the comparison between (\ref{om_low}) (say) and (\ref{om_deloc}), yields, upon using Young's equation,
 \bb
 \gamma\cos\theta_1=\gamma+\frac{8A}{L^2},
 \ee
 where we have neglected the contributions of ${\cal{O}}(L^{-3})$. For $T\lesssim T_\text{w}$ it follows that
  \bb
  -\theta_1^2=\frac{8A}{\gamma L^2}\,,
  \ee
  which can only be satisfied when the Hamaker constant is negative, i.e., only when the walls exhibit critical wetting
  (in which case the extreme of the binding potential at the centre of the pore is a local maximum).
   In this case, the critical temperature $T^*$ at
  which the system breaks the symmetry is given implicitly by
   \bb
    \theta_1(T^*)=\frac{1}{L}\sqrt{\frac{8|A(T^*)|}{\gamma(T^*)}}\,.
    \ee
   Using an abbreviation $t\equiv (T_\text{w}-T)$ and noting that $\theta(T)\sim t^{3/2}$ and $A\sim t$ for critical wetting and small $t$, we obtain an asymptotic behaviour of
   $T^*(L)$:
    \bb
    T^*=T_\text{w}-\frac{f(T^*)}{L}\,,\qquad L\to\infty, \label{cont_lim}
    \ee
where $f(T^*)>0$, in agreement with the result obtained on the basis of finite-size scaling arguments \cite{par_ev1} $T^*=T_\text{w}-L^{-1/\beta_\text{s}}$, where $\beta_\text{s}$ is the
surface critical exponent for critical wetting, for algebraically  decaying binding potential (\ref{W}) in three bulk dimensions.

\begin{figure}[!t]
\centerline{
\includegraphics[width=0.45\textwidth]{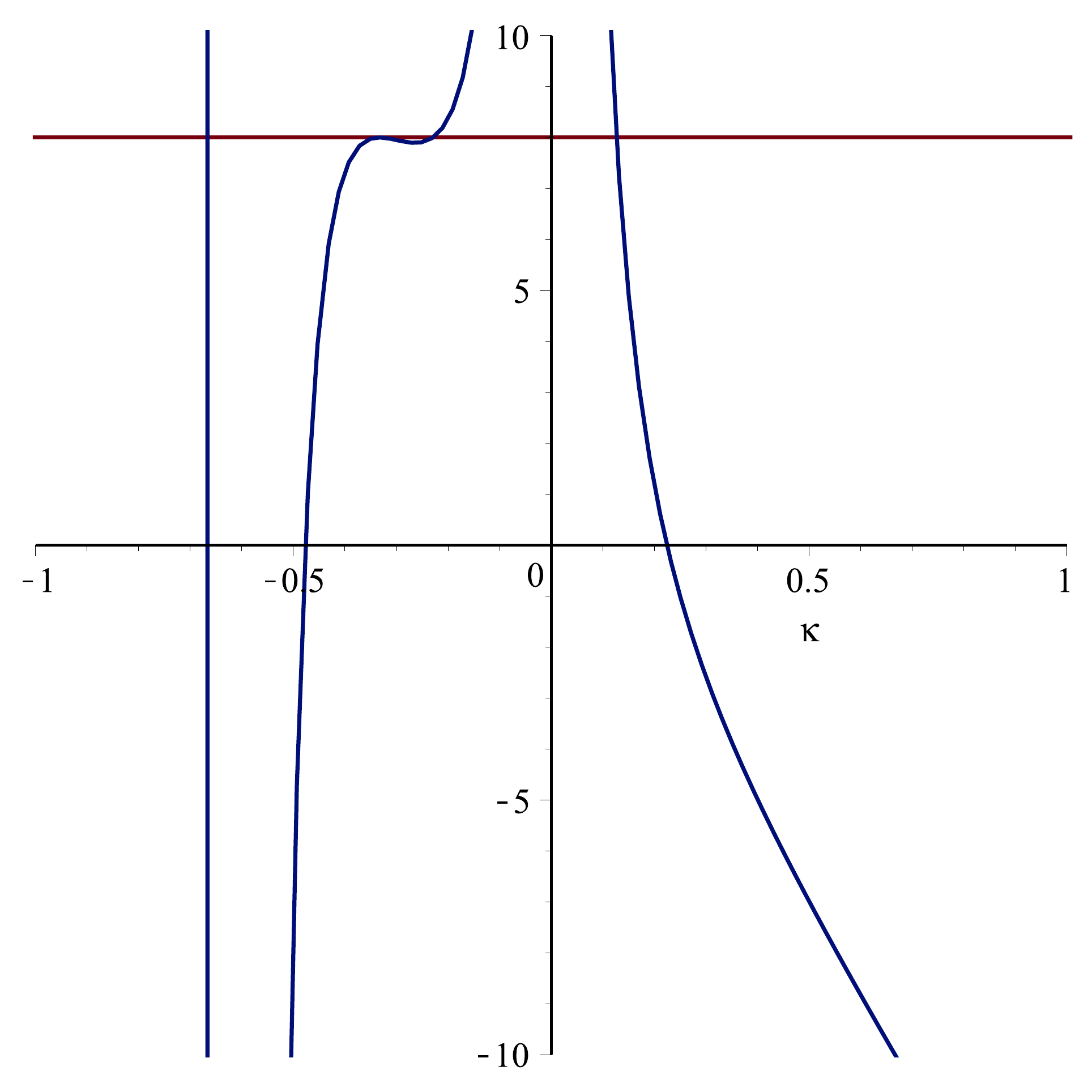}
\hspace*{0.5cm}
\includegraphics[width=0.45\textwidth]{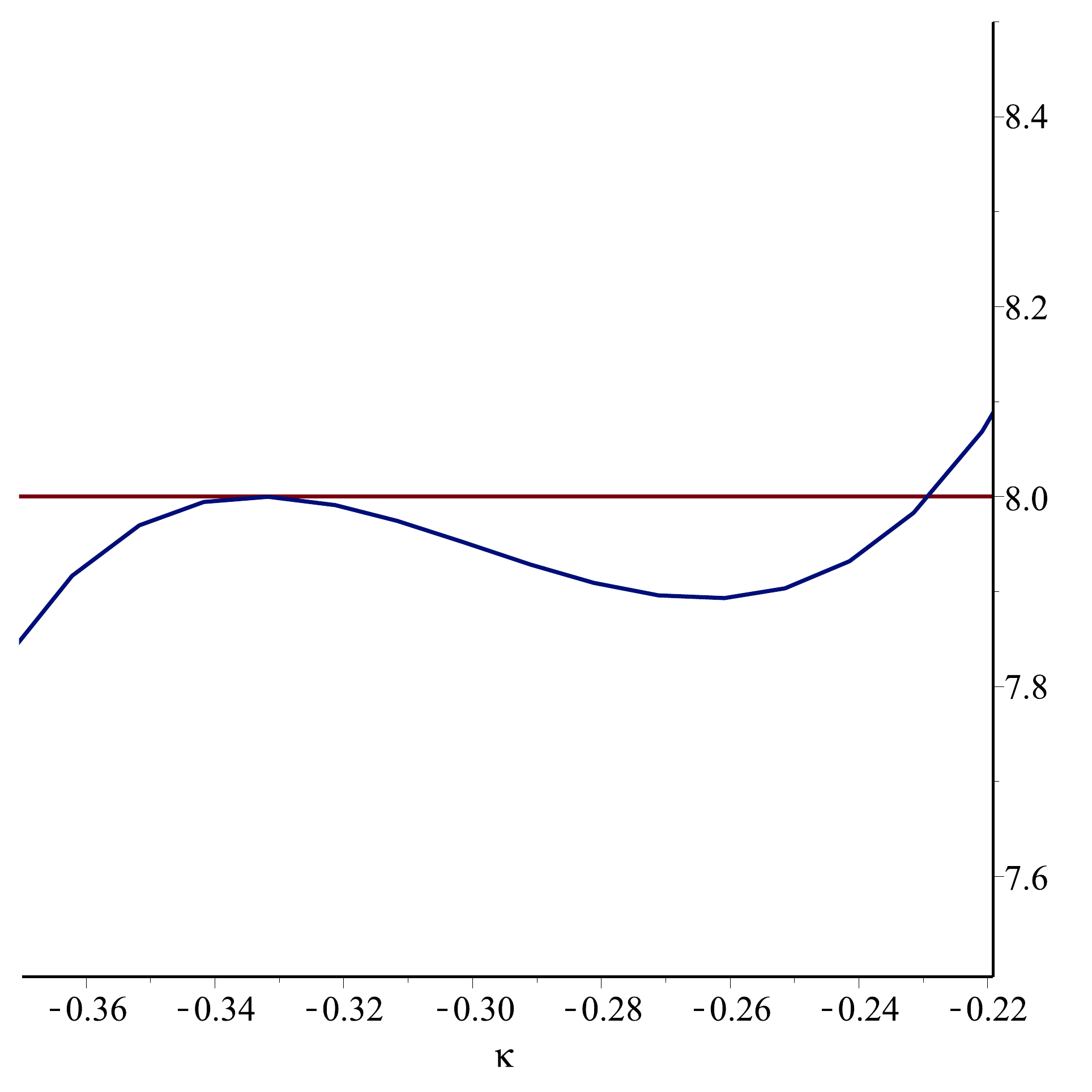}
}
\caption{A graphical solution of equation (\ref{kappa}). The curve (blue line) corresponds to the left-hand side of equation (\ref{kappa}) as a function of $\kappa=B/AL$, the
horizontal line (red) belongs to the right-hand side of equation (\ref{kappa}). The plot in the right-hand panel magnifies the graph in the interval $\kappa=-0.37$ to
$\kappa=-0.22$} \label{fig_kappa}
\end{figure}

In a more microscopic manner, the phase behaviour of a fluid confined between antisymmetric walls exhibiting critical wetting can be analyzed for $T<T_\text{w}$ by comparing the grand potential of the delocalized state
(\ref{om_deloc}) with the localized (to wall 1) one. The latter is obtained by a substitution  of $\ell=-3B/2A$ (recall that $A<0$ and $B>0$ for critical wetting below $T_\text{w}$), as given by a minimization of (\ref{W}), back to
(\ref{W}) with $\delta p=0$:
  \bb
  \omega_\text{loc}=\gamma_{\text{w}_1\text{l}}+\gamma+\gamma_{\text{w}_2\text{g}}+\frac{4}{27}\frac{A^3}{B^2}+\frac{A}{\left(L+\frac{3}{2}\frac{B}{A}\right)^2}
                    +\frac{B}{\left(L+\frac{3}{2}\frac{B}{A}\right)^3}\,.
  \ee
   Balancing $\omega_\text{loc}$ and $\omega_\text{deloc}$ leads to the equation
   \bb
   \frac{4}{27}\frac{1}{\kappa^2}+\frac{1}{\left(1+\frac{3}{2}\kappa\right)^2}
   +\frac{\kappa}{\left(1+\frac{3}{2}\kappa\right)^3}-16\kappa=8,\label{kappa}
   \ee
where we have introduced a dimensionless parameter $\kappa\equiv B/AL$. Equation (\ref{kappa}) can be solved graphically  as is shown in figure~\ref{fig_kappa} revealing four solutions with three of them, corresponding to
negative $\kappa$, being relevant. This analysis suggests that within the interval between $\kappa=-2/3$, where the LHS of equation (\ref{kappa}) has a pole and $\kappa\approx-0.23$, the delocalized state is the stable solution,
while the localized state is the more stable solution otherwise. There is another point $\kappa=-1/3$ at which the localized and delocalized states are equally stable; however, this solution seems to be just an artifact of the
current analysis which presumably disappears when higher order terms of ${\cal{O}}(L^{-4})$ in the binding potential are taken into account. We note that this result is not inconsistent with the asymptotic result given  by
equation (\ref{cont_lim}),  since in the latter case $L$ is large and $A$ is small, so that $|\kappa|$ is not necessarily small in this limit.

For $T>T_\text{w}$, there is only a delocalized state present for the walls undergoing critical wetting. For the walls exhibiting first-order wetting transition, the localized states become metastable with respect to the delocalized
one in the macroscopic limit of the separation of the walls. For $L$ finite, the binding potential is non-vanishing in the middle of the pore which makes the delocalized state less stable. The grand potential per unit area for the
localized state is given by
 \bb
  \omega_\text{loc}=\gamma_{\text{w}_1\text{l}}+\gamma+\gamma_{\text{w}_2\text{g}}+W_1(\ell_\text{loc})+W_2(L-\ell_\text{loc}) , \label{loc2}
  \ee
where $\ell_\text{loc}$ is the minimum of the binding potential near the (left-hand) wall. In order to describe the binding potential for the case of first-order wetting (see figure~\ref{bp_sketch}), we should consider one more term in~(\ref{W})
  \bb
  W_i(\ell)=\frac{A_i}{\ell^2}+\frac{B_i}{\ell^3}+\frac{C_i}{\ell^4}+\cdots, \label{W2}
  \ee
where $A_i$, $C_i>0$ and $B_i<0$ and where we set $\delta p=0$. Minimization of $ W_i(\ell)$ gives $\ell_\text{loc}=-3B/4A-\sqrt{9B^2/A^2-32C/A}/4$, where we again assume
that the walls are perfectly antisymmetric ($A_1=A_2$, $B_1=B_2$ and $C_1=C_2$ and thus $W_1=W_2\equiv W$). The grand potential per unit area for the delocalized state
is as follows:
 \bb
  \omega_\text{deloc}=\gamma_{\text{w}_1\text{l}}+\gamma+\gamma_{\text{w}_2\text{g}}+2W(L/2)\,. \label{deloc2}
  \ee

The energy barrier of $W(\ell)$ disappears when its local minimum and maximum coincide which leads to the condition for the spinodal temperature $T_\text{s}$:
 \bb
 \frac{B^2}{AC}=\frac{32}{9}\,,\qquad T=T_\text{s}\,.
 \ee
Above $T_\text{s}$, only the delocalized state is present. For $T_\text{w}<T<T_\text{s}$,  there exists a wall separation $L^*(T)$ for which $\omega_\text{loc}=\omega_\text{deloc}$, which
determines the location of the interface delocalization transition. At this wall separation, the presence of the walls rises (relative to the case of $L$ macroscopic)
the central minimum of the binding potential (corresponding to the delocalized state) up to the level of the minima near the walls (corresponding to the localized
states).
 For $T\approx T_\text{s}$ we  approximately obtain
  \bb
  L_\text{c}\equiv L^*(T_\text{s})\approx\frac{3}{\sqrt{2}}\frac{|B|}{A}\,,\label{lc}
  \ee
which is the critical value of $L$ below which the interface localized-delocalized transition never occurs.  In the opposite limit, $L\to\infty$, the transition occurs
right at $T_\text{w}$ in line with the macroscopic arguments. Below $T_\text{w}$, only the states with the interface pinned to either walls are stable for any value of $L$.

Finally, we note that in a more general case of competing walls when $A_1\neq A_2$, the position of the delocalized interface is shifted towards the wall with a weaker
surface field, such that
 \bb
 \ell_\text{eq}\approx\frac{L}{\left(1+\sqrt[3]{A_2/A_1}\right)}\,.
 \ee

\section{Density functional theory}

\label{sec:3}

In order o describe microscopic properties of a simple fluid in heterogeneous pores we adopt a classical density function theory \cite{evans79}. Within DFT, the
equilibrium density profile is found by minimizing the grand potential functional
 \bb
 \Omega[\rho]={\cal{F}}[\rho]+\int\dr\left[V(\rr)-\mu\right]\rhor\,,\label{om}
 \ee
where $V(\rr)$ is the external field due to the walls and $\mu$ is the chemical potential. The intrinsic free energy functional ${\cal{F}}$ can be split into the contribution from the ideal gas and the remaining excess part
 \bb
 {\cal{F}}[\rho]={\cal{F}}_{\rm id}[\rho]+{\cal{F}}_{\rm ex}[\rho]\,,
 \ee
where ${\cal{F}}_{\rm id}[\rho]= k_\text{B}T\int\dr\rhor\left[\ln(\Lambda^3\rhor)-1\right]$ with $\Lambda$ being the thermal de Broglie wavelength that can be set to unity.

The fluid model is characterised by a pair potential consisting of a hard-sphere repulsion of the range of $\sigma$, and a Lennard-Jones-like attractive portion given by
  \bb
 u(r)=\left\{\begin{array}{ll}
 0\,,&r<\sigma,\\
 -4\varepsilon\left(\frac{\sigma}{r}\right)^6,& \sigma<r<r_\text{c},\\ \label{u}
 0\,,&r\geqslant r_\text{c},
\end{array}\right.
 \ee
 where the cut-off is set to $r_\text{c}=2.5\,\sigma$. The repulsive contribution to the excess free energy functional is approximated using Rosenfeld's  fundamental measure theory \cite{ros} and the attractive part is
 treated within the mean-field approximation:
 \bb
{\cal{F}}_{\rm ex}[\rho]=k_\text{B}T\int\dr\,\Phi(\{n_\alpha\})+\frac{1}{2}\int\dr\rhor\int\dr'\rho(\rr')u(|\rr-\rr'|)\,,\label{fex}
 \ee
The free energy density $\Phi$ is a function of a set of three independent weighted densities $\{n_\alpha\}$ for which we use the original Rosenfeld prescription.

The external field due to the parallel impenetrable walls a distance $L$ apart is given by $V(\rr)=V(z)=V_1(z)+V_2(L-z)$, where
  \bb
 V_i(z)=\left\{\begin{array}{ll}
 \infty\,,& z<\sigma\,,\\
 -\frac{2}{3}\frac{\pi\varepsilon_{\text{w}_i}\rho_{\text{w}_i}\sigma^6}{z^3}\,,&z\geqslant \sigma\,.
\end{array}\right. \label{pot_wall}
 \ee
The parameter $\rho_{\text{w}_i}$ is a density of uniformly distributed atoms forming the wall $i$, such that each atom exerts a potential  $u(r)$ according to expression
(\ref{u}) with a parameter $\varepsilon_{\text{w}_i}$ replacing $\varepsilon$ and without a cut-off, i.e., for $r_\text{c}=\infty$. The potential $V_i$ of a single wall $i$  is given
by integrating all contributions of atoms over the entire wall. The diameter $\sigma$ of the wall atoms has been chosen equal to that of fluid atoms. Since the parameter
$\rho_{\text{w}_i}$ is always associated with $\varepsilon_{\text{w}_i}$, we can characterise the wall strength by tuning just a single parameter $\varepsilon_{\text{w}_i}$ by setting
$\rho_{\text{w}_1}=\rho_{\text{w}_2}=\sigma^{-3}$.

In order to make a link with the more phenomenological model introduced in the previous section,  the Hamaker constant defined by equation (\ref{W}) can be expressed in terms of the microscopic parameters using the sharp-kink
approximation \cite{dietrich}. For the interaction between wall-liquid  and liquid-gas interfaces we have
 \begin{eqnarray}
 A_i=\frac{1}{3}\pi\varepsilon_{\text{w}_i}\rho_{\text{w}_i}(\rho_\text{l}-\rho_\text{g})\sigma^6\,, \label{A}\\
 B_i=-\frac{2}{3}\pi\varepsilon_{\text{w}_i}\rho_{\text{w}_i}(\rho_\text{l}-\rho_\text{g})\sigma^7\,, \label{B}
 \end{eqnarray}
where it should be noted that $A_i$ is always positive in our model owing to the fact that the fluid-fluid interaction (\ref{u}) is only short ranged and thus does not
contribute to $A_i$. For the interaction between wall-gas and gas-liquid interfaces, the expressions in equations~(\ref{A}) and (\ref{B}) have opposite signs. However, for
``antisymmetric'' walls, $\varepsilon_{\text{w}_1}=-\varepsilon_{\text{w}_2}$, in which case $A_1=A_2$ and $B_1=B_2$. This ``antisymmetry'' is deceptive, however, even if
$\varepsilon_{\text{w}_1}=-\varepsilon_{\text{w}_2}$ one does not expect $B_1$ and $B_2$ to be identical beyond the sharp-kink approximation, since these second-order contributions
are due to excluded volume effects at the walls and it is reasonable to expect that those of the wall-liquid interface are much stronger than those of the wall-gas
interface. Indeed, one does not expect the existence of a drying temperature at all for the purely repulsive wall, so that a link between a microscopic and a  mesoscopic
model at the level of the sharp-kink approximation is fully justified only up to the leading order with a coefficient $A_i$.

\begin{figure}[!b]
\centerline{
\includegraphics[width=0.55\textwidth]{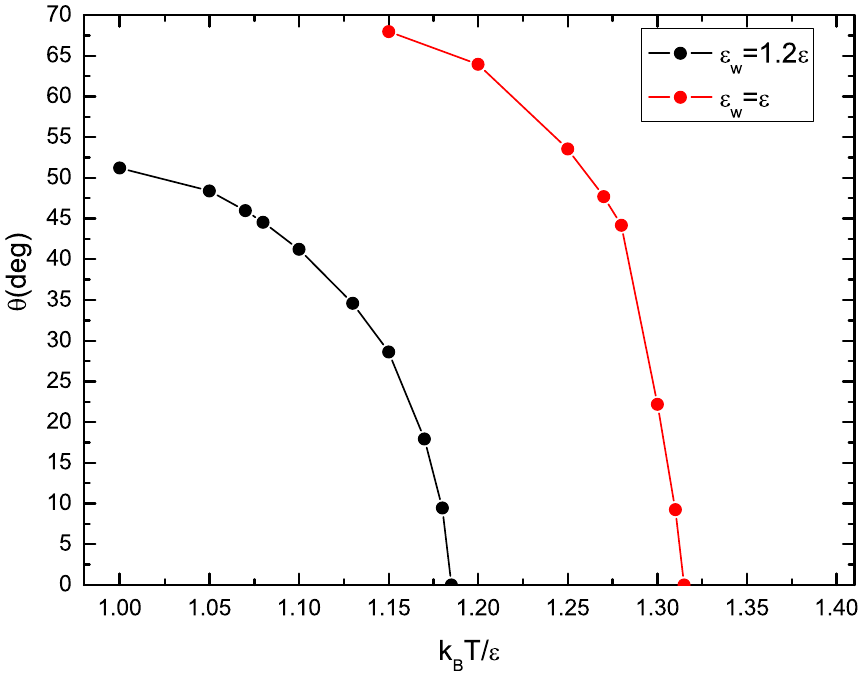}
}
\vspace{-2mm}
\caption{(Color online) Variation of the contact angle with temperature for walls with strengths  $\varepsilon_{\text{w}_1}=1.2\,\varepsilon$ and $\varepsilon_{\text{w}_2}=\varepsilon$.} \label{cont_ang}
\end{figure}

The minimization of (\ref{om}) is carried out numerically on a 2D grid with spacing $\dd x=\dd z=0.05\,\sigma$. Although the symmetry of the external field permits to
treat the problem as one-dimensional $\rhor=\rho(z)$, we also wish to test the geometric arguments for microscopic values of $L$ by constructing equilibrium density
profiles of a single pore filled with liquid- and gas-like coexisting phases, in which case $\rhor=\rho(x,z)$. For a given set of model and thermodynamic parameters, the
1D DFT results are used as boundary conditions for the 2D DFT, such that we set $\rho_{2\text{D}}(x,z_{\rm max})=\rho^{\rm low}_{1\text{D}}(x)$  and $\rho_{2\text{D}}(x,z_{\rm
min})=\rho^{\rm high}_{1\text{D}}(x)$. Here, $z_{\rm max}$ and $z_{\rm min}$ are respectively the maximal and minimal values of the vertical coordinate $z$ which are considered
within the 2D calculations and are set to  $z_{\rm min}=0$ and $z_{\rm max}=40\,\sigma$. The functions $\rho^{\rm low}_{1\text{D}}(x)$ and $\rho^{\rm high}_{1\text{D}}(x)$ are 1D
density profiles corresponding to the low and high density states, respectively. The low density state is a configuration in which the system is filled primarily with a
gas (the liquid-gas interface is pinned to the left-hand wall), whereas the high density state is a configuration in which the system is either filled with liquid (the
interface is pinned to the right-hand wall) or a delocalized state in which case the liquid-gas interface is around the centre of the pore. Further details and particularly
the implementation of Rosenfeld's functional within the 2D-DFT treatment can be found in reference~\cite{mal_groove}.

\begin{figure}[!t]
\centerline{
\includegraphics[height=6cm]{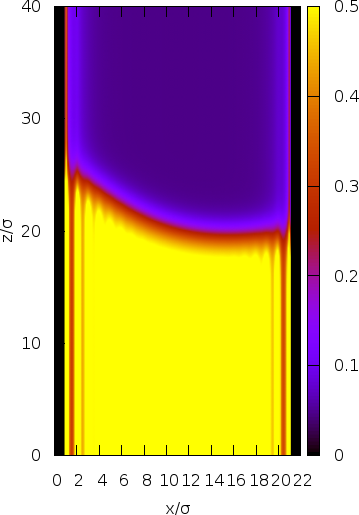}
\hspace*{0.5cm}
\includegraphics[height=6cm]{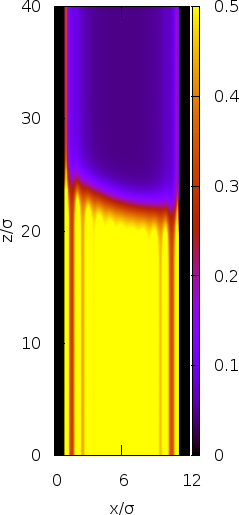}
\hspace*{0.5cm}
\includegraphics[height=6cm]{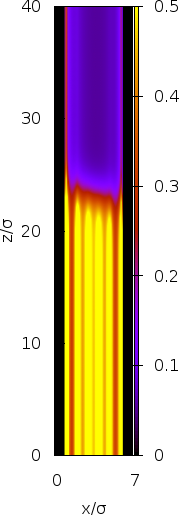}
}
\caption{(Color online) Two-dimensional equilibrium density profiles of a fluid confined between walls  with surface fields $\varepsilon_{\text{w}_1}=1.2\,\varepsilon$ and
$\varepsilon_{\text{w}_2}=\varepsilon$ at a temperature $T=0.81\,T_\text{c}$ ($k_\text{B}T/\varepsilon=1.15$) which is below both $T_{\text{w}_1}$ and $T_{\text{w}_2}$. The contact angles at the single
walls are $\theta_1=29\degree$ and $\theta_2=68\degree$, as given by the results displayed in figure~\ref{cont_ang}. The width of the pores accessible for fluid particles
is (from left to right) $L=20\,\sigma$, $L=10\,\sigma$, and $L=5\,\sigma$. All configurations correspond to the liquid-gas coexistence in the respective pores, i.e., the
capillary condensation which occurs at the chemical potentials that are shifted below $\mu_\text{sat}=-4.0309\,\varepsilon$ by the values
$\delta\mu(L=20\,\sigma)=0.0275\,\varepsilon$, $\delta\mu(L=10\,\sigma)=0.0547\,\varepsilon$ and $\delta\mu(L=5\,\sigma)=0.1934\,\varepsilon$.} \label{T115}
\end{figure}

We start the discussion of our numerical DFT results by examining the wetting properties of a single wall. In figure~\ref{cont_ang} we display the temperature dependence of
the contact angle for two walls with surface fields $\varepsilon_{\text{w}_1}=1.2\,\varepsilon$ (wall 1) and $\varepsilon_{\text{w}_2}=\varepsilon$ (wall 2). The walls exhibit
first-order wetting transition at temperatures $T_{\text{w}_1}=0.83\,T_\text{c}$ and $T_{\text{w}_2}=0.93\,T_\text{c}$, with $k_\text{B}T_\text{c}/\varepsilon=1.414$ corresponding to the bulk critical
temperature \cite{wedge_prl}. We now consider a pore formed by a wall 1 and a wall 2 at a temperature  $T=0.81\,T_\text{c}$ $(k_\text{B}T/\varepsilon=1.15)$ at which both walls, when
separated, are only partially wet. In figure~\ref{T115} we display the equilibrium 2D density profiles corresponding to the liquid-gas coexistence in pores of widths
$L=20\,\sigma$, $L=10\,\sigma$, and $L=5\,\sigma$. In this case, the walls of the pore in the low-density phase are only microscopically wet and the liquid-gas meniscus
meets both walls at angles that appear in a  good agreement with the predicted values of the respective contact angles (cf. figure~\ref{cont_ang}), including, somewhat
surprisingly, even the narrow pore with $L=5\,\sigma$. We have further determined the location of capillary condensation for each of the pores, and the results are in
perfect agreement with the modified Kelvin's equation (\ref{kelvin_het}), cf. table~\ref{tab:dmu}.

\begin{table}[!b]
 \begin{center}
 \caption{The predicted values of a location of capillary condensation, as given by Kelvin's equation (\ref{kelvin_het}) and its modified version taking into account the
  effect of wetting layers [equation (\ref{kelvin_het2})], are compared with DFT results
  for three pore widths and temperatures $T<T_{\text{w}_1}<T_{\text{w}_2}$, $T_{\text{w}_1}<T<T_{\text{w}_2}$ and $T_{\text{w}_1}<T_{\text{w}_2}<T$. The comparison is made for the heterogeneous pore with
   $\varepsilon_{\text{w}_1}=1.2\,\varepsilon$ and $\varepsilon_{\text{w}_2}=\varepsilon$.}
 \vspace*{2mm}
 \errorcontextlines 1000
 \begin{tabular}{|c|l||c| c| c|}
 \hline\hline
Temperature &Pore width &DFT& equation~(\ref{kelvin_het})& equation~(\ref{kelvin_het2})\\
  \hline\hline
 &$L=20\,\sigma$ & $\delta\mu=0.027502\,\varepsilon$  & $\delta\mu=0.026998\,\varepsilon$ & --  \\

$k_\text{B}T/\varepsilon=1.15$  &$L=10\,\sigma$ & $\delta\mu=0.054670\,\varepsilon$  & $\delta\mu=0.053996\,\varepsilon$ & --  \\

  &$L=5\,\sigma$ & $\delta\mu=0.101590\,\varepsilon$  & $\delta\mu=0.107991\,\varepsilon$ & --  \\
  \hline
 &$L=20\,\sigma$ & $\delta\mu=0.027415\,\varepsilon$  & $\delta\mu=0.020330\,\varepsilon$ & $\delta\mu=0.026229\,\varepsilon$  \\

$k_\text{B}T/\varepsilon=1.3$  &$L=10\,\sigma$ & $\delta\mu=0.059021\,\varepsilon$  & $\delta\mu=0.040635\,\varepsilon$ & $\delta\mu=0.058079\,\varepsilon$  \\

  &$L=5\,\sigma$ & $\delta\mu=0.121618\,\varepsilon$  & $\delta\mu=0.081302\,\varepsilon$ & $\delta\mu=0.116158\,\varepsilon$  \\
  \hline
$k_\text{B}T/\varepsilon=1.35$ &$L=20\,\sigma$ & $\delta\mu=0.0234047\,\varepsilon$  & $\delta\mu=0.014348\,\varepsilon$ & $\delta\mu=0.026087\,\varepsilon$  \\
  \hline\hline
 \end{tabular}
 \label{tab:dmu}
 \end{center}
\end{table}

\begin{figure}[!t]
\centerline{
\includegraphics[height=6cm]{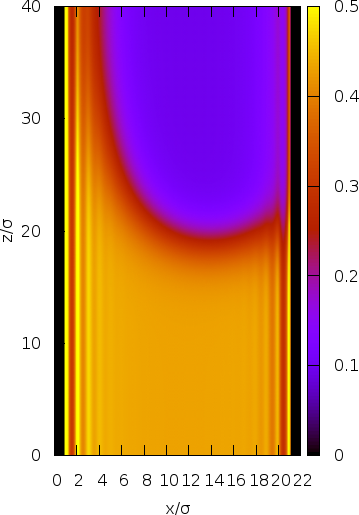}
\hspace*{0.5cm}
\includegraphics[height=6cm]{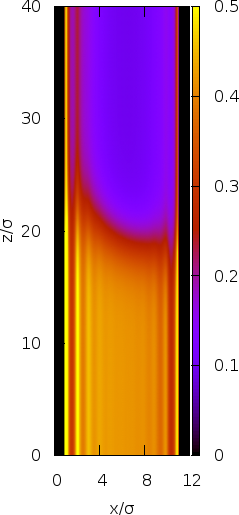}
\hspace*{0.5cm}
\includegraphics[height=6cm]{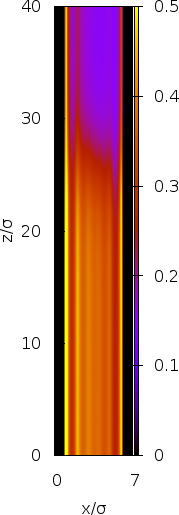}
}
\caption{Two-dimensional equilibrium density profiles of a fluid confined between walls  with surface fields $\varepsilon_{\text{w}_1}=1.2\,\varepsilon$ and
$\varepsilon_{\text{w}_2}=\varepsilon$ at a temperature $T=0.92\,T_\text{c}$ ($k_\text{B}T/\varepsilon=1.3$) which is below $T_{\text{w}_2}$ but above $T_{\text{w}_1}$. The contact angle of the partially
wet wall is $\theta_1=22\degree$, as given by the results displayed in figure~\ref{cont_ang}. The width of the pores accessible for fluid particles is (from left to right)
$L=20\,\sigma$, $L=10\,\sigma$, and $L=5\,\sigma$. All configurations correspond to the liquid-gas coexistence in the respective pores, i.e., the capillary condensation
which occurs at the chemical potentials that are shifted below $\mu_\text{sat}=-3.9651\,\varepsilon$ by the values $\delta\mu(L=20\,\sigma)=0.0275\,\varepsilon$,
$\delta\mu(L=10\,\sigma)=0.0590\,\varepsilon$ and $\delta\mu(L=5\,\sigma)=0.1216\,\varepsilon$.} \label{T13}
\end{figure}

\begin{figure}[!b]
\centerline{
\includegraphics[height=6cm]{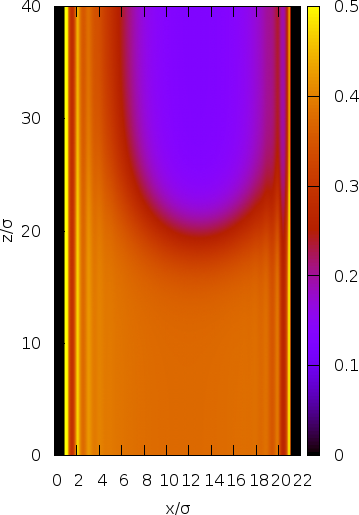}
}
\caption{Two-dimensional equilibrium density profile of a fluid confined between walls  with surface fields  $\varepsilon_{\text{w}_1}=1.2\,\varepsilon$ and
$\varepsilon_{\text{w}_2}=\varepsilon$ at a temperature $T=0.95\,T_\text{c}$ ($k_\text{B}T/\varepsilon=1.35$) which is above both $T_{\text{w}_1}$ and $T_{\text{w}_2}$. The width of the pores accessible
for fluid particles is $L=20\,\sigma$ and the configuration corresponds to the liquid-gas coexistence and occurs at the chemical potentials that is shifted below $\mu_\text{sat}=-3.9569\,\varepsilon$ by the value $\delta\mu(L=20\,\sigma)=0.0234\,\varepsilon$.} \label{T135}
\end{figure}

We next consider a temperature $T=0.92\,T_\text{c}$ $(k_\text{B}T/\varepsilon=1.3)$, in which case the wall 2 is still partially wet but the wall 1, when separated and at bulk two-phase coexistence, would be completely wet. The equilibrium
density profiles at two-phase coexistence are shown in figure~\ref{T13}. Now, the comparison of the DFT results with the macroscopic Kelvin equation (\ref{kelvin_het}) is less satisfactory than in the previous case, where both
walls are non-wet, especially for narrow pores, see table~\ref{tab:dmu}. Nevertheless, if we employ the modified Kelvin equation as given by (\ref{kelvin_het2}), we obtain a prediction in a very good agreement with DFT,
provided the capillaries are sufficiently wide. For instance, for $L=50\,\sigma$, the relative difference in $\delta\mu$ between DFT and the modified Kelvin equation is only $4\%$ (compared to $18\%$ for the macroscopic Kelvin
equation). Only for those large capillaries, the assumptions leading to equation~(\ref{kelvin_het2}) are justified.  For the pores widths considered here, one can still use equation~(\ref{kelvin_het2}) but with $\ell_\pi$ replaced by
the real thickness of the wetting layer. This can be read off from the 1D density profiles by determining e.g., the Gibbs dividing surface of the liquid-gas interface. The inclusion of this correction substantially improves the
agreement with the DFT results, as is shown in table~\ref{tab:dmu}.

Finally, if we increase the temperature above $T_{\text{w}_2}$, then both walls are covered with wetting films but of different thickness, as can be seen from the equilibrium
density profile for $L=20\,\sigma$ and $\mu=\mu_\text{cc}$ in figure~\ref{T135}. As expected, the presence of the wetting layers deteriorates the quality of the prediction
given by the macroscopic Kelvin equation [equation~(\ref{kelvin_het})]  but the modified Kelvin equation [equation~(\ref{kelvin_het2})] is still fairly accurate, when again both
$\ell_\pi^{(i)}$ are replaced by the respective film thicknesses, see table~\ref{tab:dmu}. At this temperature ($T=0.95\,T_\text{c}$), there is no capillary condensation for
the intermediate ($L=10\,\sigma$) and narrow ($L=5\,\sigma$) pores. While the $\mu-\Omega$ dependence (not shown here) is smooth for the narrowest pore. It exhibits a kink
for the intermediate pore suggesting that the critical pore width at this temperature is~$L_\text{c}\approx10\,\sigma$.

\begin{figure}[!b]
\centerline{
\includegraphics[width=9cm]{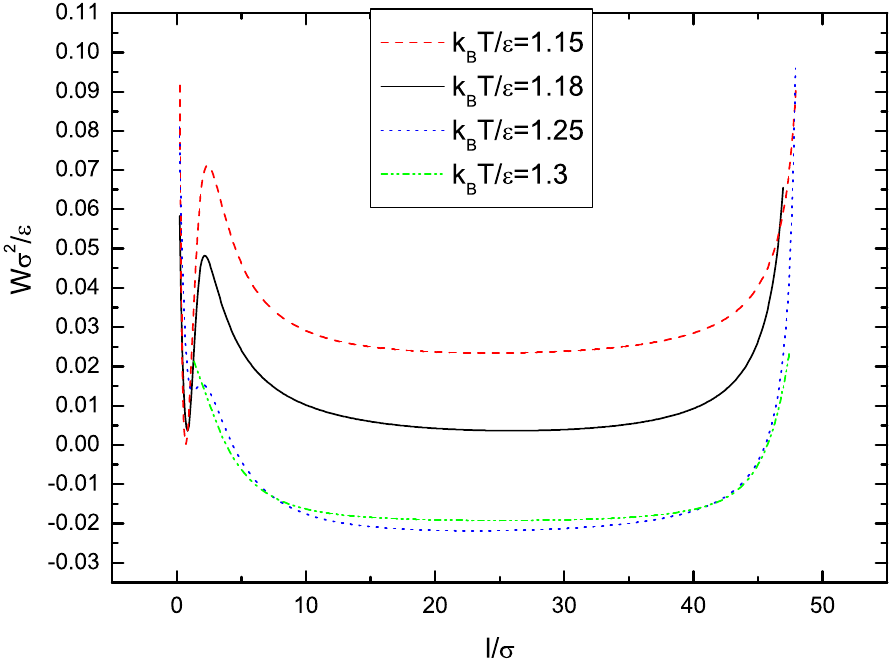}
}
\caption{Binding potentials in a capillary with competing walls ($\varepsilon_{\text{w}_1}=-\varepsilon_{\text{w}_2}=1.2\,\varepsilon$) and a width $L=50\,\sigma$.  For all
temperatures below the spinodal temperature $T_\text{s}$ the binding potential exhibits two minima, corresponding to the liquid-gas interface being pinned at wall 1 and a
delocalized interface with a mean location in the pore midpoint. The competition between the two minima determines the equilibrium configuration of the fluid. For this
wide pore, the crossover temperature $T^*$ between the states, at which  the interface is bounded to the wall ($T<T^*$) and delocalized  ($T>T^*$) is $T^*\approx T_\text{w}$
($k_\text{B}T_\text{w}/\varepsilon=1.18$ for wall 1). The highest temperature ($k_\text{B}T=1.3\,\varepsilon$) is already above $T_\text{s}$ and thus the binding potential has only one minimum in
the middle of the pore. Also note a very small energetic barrier for $k_\text{B}T=1.25\,\varepsilon$.} \label{bp50}
\end{figure}

At last, we turn our attention to the case of asymmetric walls, such that one wall tends to be wet and the other dry. To this end, we consider a capillary in which the
wall 1 has the same strength as before, i.e., $\varepsilon_{\text{w}_1}=1.2\,\varepsilon$ but the opposite wall is purely repulsive. The repulsive wall exerts the potential
according to (\ref{pot_wall})   with a \emph{negative} surface field $\varepsilon_{\text{w}_2}=-1.2\,\varepsilon$. Although the Hamaker
constants and thus the asymptotic behaviour of binding potentials for both walls are assumed to be the same, it is only the attractive wall (wall 1) which exhibits (first
order) wetting transition at temperature $T_\text{w}$. The purely repulsive wall (wall 2), does not induce a drying transition, i.e., the wall is always completely wet by gas.
Consequently, the binding potential  of wall 2 is purely repulsive (monotonously decaying) in contrast with the binding potential of wall 1 which, at least for
sufficiently low temperatures, has two competing minima that are of equal depth at $T=T_\text{w}$. In figure~\ref{bp50} we display numerically constructed binding potentials for
this model pore using a constrained minimization of (\ref{om}) at bulk two-phase coexistence for several representative temperatures and a relatively wide pore
($L=50\,\sigma$). For $T<T_\text{w}$, the global minimum in the total binding potential lies near wall 1, which means that the liquid-gas interface is pinned to the adsorbing
wall and thus the system is in a low-density state for $\mu\leqslant\mu_\text{sat}$. On the other hand, if the two-phase coexistence is approached from the liquid state, the
pore exhibits capillary emptying for some $\mu>\mu_\text{sat}$, according to the Kelvin equation (\ref{kelvin_het}). For $T\approx T_\text{w}$, the binding potential has two
equally deep minima: one near the wall 1 and the other at the middle of the pore. For a perfectly antisymmetric system (such as for magnets), this temperature would be
identified with a triple point, where the states with a liquid-gas interface near both walls and the state with the liquid-gas interface in the centre of the pore are all
equally stable. In this model, however, the configuration corresponding to the pore filled with liquid is missing (at $\mu=\mu_\text{sat}$), so that the transition at a
temperature $T^*\approx T_\text{w}$ can be interpreted as a thin-to-thick transition in some analogy with prewetting transition on a single wall. However, at a single wall, the
prewetting transition is induced by the appearance of a non-zero term $\delta p\ell$ in the binding potential (\ref{W}), i.e., because the system is away from bulk
coexistence, which shifts a local minimum of $W(\ell)$ at $\ell=\infty$ (at saturation, $\delta p=0$) to a finite distance from the wall (for $\delta p>0$). For our
system, the shift of the minimum is due to the presence of the second wall. For a temperature $T>T_\text{w}$, the minimum of $W(\ell)$ at $\ell=L/2$ becomes a global minimum
corresponding to a delocalized interface. Above the spinodal temperature $T_\text{s}$, the energy barrier disappears and the configuration with the bounded interface ceases to
be even metastable.

\begin{figure}[!t]
\centerline{
\includegraphics[width=9cm]{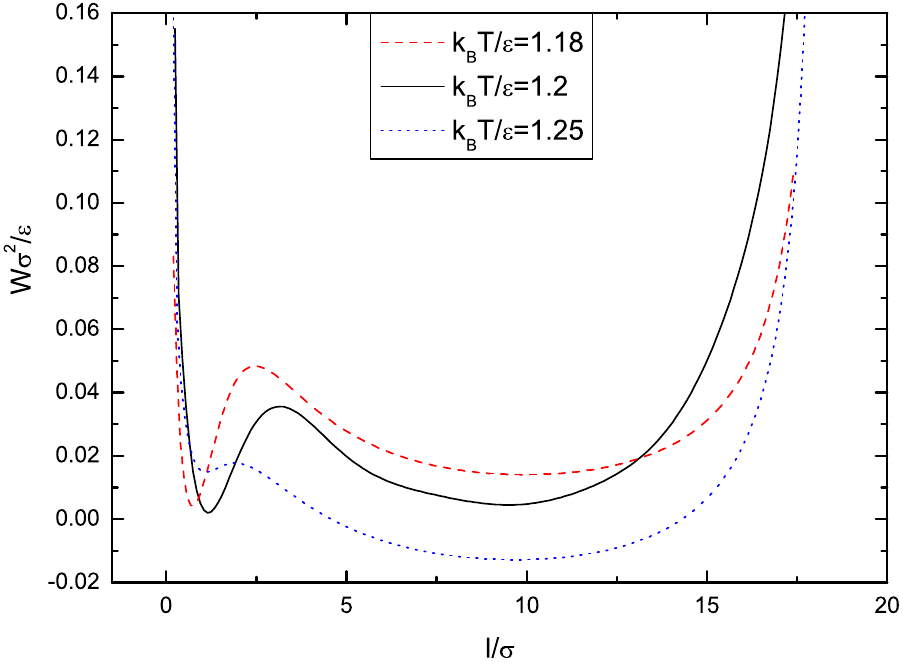}
}
\caption{Binding potentials in a capillary with competing walls ($\varepsilon_{\text{w}_1}=-\varepsilon_{\text{w}_2}=1.2\,\varepsilon$) and a width $L=20\,\sigma$. In this case, in
contrast to the case of a wide pore shown if figure~\ref{bp50}, the temperature $T^*$ at which two different fluid configurations coexist is above $T_\text{w}$ and corresponds to
$k_\text{B}T\approx1.2\,\varepsilon$.} \label{bp20}
\end{figure}

\begin{figure}[!b]
\centerline{
\includegraphics[height=6cm]{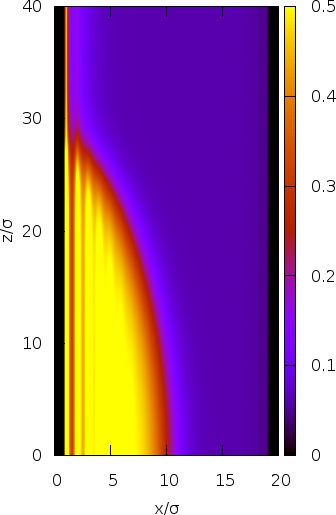}
}
\caption{Two-dimensional density profile of a fluid confined between parallel walls with strengths $\varepsilon_\text{w}=1.2\,\varepsilon$  (left-hand wall) and
$\varepsilon_\text{w}=-1.2\,\varepsilon$ (right-hand wall), separated by a distance $L=20\,\sigma$. The temperature of the system $k_\text{B}T=1.2\,\varepsilon$ corresponds to the
interface localized-delocalized transition ($\mu=\mu_\text{sat}$).} \label{p_antisym}
\end{figure}

\begin{figure}[!t]
\centerline{
\includegraphics[width=9cm]{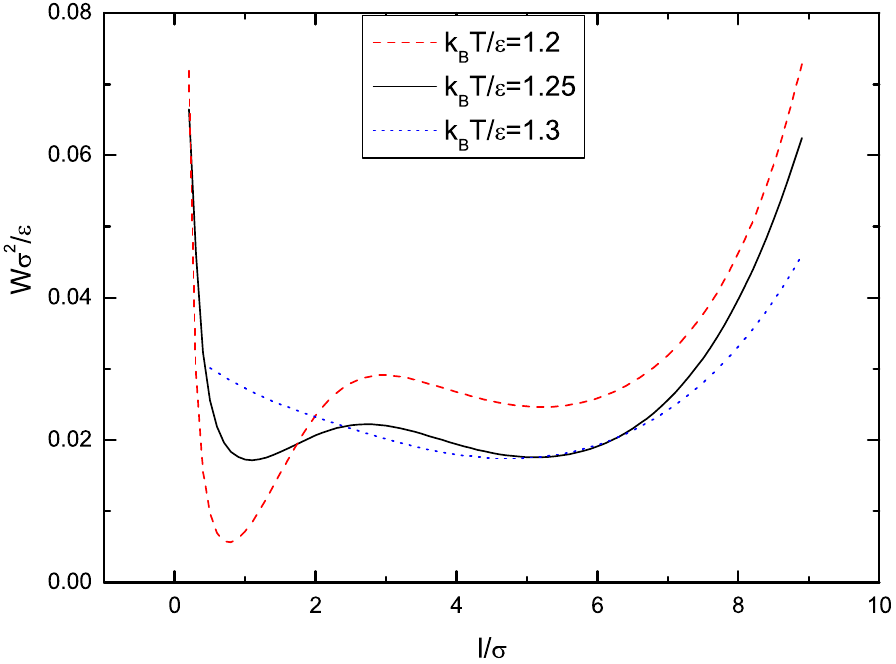}
}
\caption{Binding potentials in a capillary with competing walls ($\varepsilon_{\text{w}_1}=-\varepsilon_{\text{w}_2}=1.2\,\varepsilon$) and a width $L=10\,\sigma$. The temperature $T^*$ at which two different fluid configurations coexist is
now $k_\text{B}T\approx1.25\,\varepsilon$, which is only slightly below the spinodal temperature (cf. figure~\ref{bp50}). For $T>T_s$, the binding potential has only one minimum in the pore midpoint.} \label{bp10}
\end{figure}

Thus far, the DFT results are in line with the predictions obtained from macroscopic considerations. However, if the pore width is reduced, the midpoint minimum of the binding potential becomes more affected by the (repulsive)
interaction between the liquid-gas interface with both walls. Consequently, the midpoint minimum is pushed upwards which brings about a two-phase coexistence at $T^*>T_\text{w}$, as illustrated in figure~\ref{bp20} for the pore width
of $L=20\,\sigma$. In figure~\ref{p_antisym} we also show a density profile for a temperature $k_\text{B}T=1.2\,\varepsilon$ at which the states corresponding to the localized and delocalized interface coexist.  As we reduce the pore
width even more, $T^*(L)$ still increases and terminates at $T_\text{s}$ which determines the critical pore width $L_\text{c}=L(T^*=T_s)$ below which no phase transition occurs at $\mu=\mu_\text{sat}$. In figure~\ref{bp10} we display binding
potentials for a near critical pore width ($L=10\,\sigma$); note that the two-phase coexistence occurs now at a temperature $k_\text{B}T/\varepsilon=1.25$ for which the binding potential for a single wall (or a wide pore) has only a
very weak minimum near the wall (cf. figure~\ref{bp50}). Also shown here is the binding potential for $T>T_s$ which does not permit any phase transition at $\mu=\mu_\text{sat}$ for any pore width.

\section{Summary and concluding remarks}

\label{sec:4}

In this work we have studied the phase behaviour of a simple fluid confined between two parallel walls with different surface fields in the presence of long-range dispersion
forces. We have shown that simple geometric arguments lead to the generalized macroscopic Kelvin equation which, when at least one of the walls is in complete wetting
regime, should be modified by capturing the effect of thick wetting films by including Derjaguin's correction. We have also shown that simple mean-field arguments can be
invoked to describe the fluid phase behaviour in a capillary at which one wall tends to be wet while the other tends to be dry and which exhibits interface
localized-delocalized transition. Some of these predictions have been tested using a microscopic non-local density functional theory. The main results of this work can
be summarised as follows:

\begin{enumerate}

\item Simple geometric arguments lead to a macroscopic prediction for a location (a shift in the chemical potential relative to that of bulk saturated fluid)
of the capillary condensation/evaporation in a pore made of different walls. This leads to the (generalized) Kelvin equation (\ref{kelvin_het}) which has been tested
against DFT calculations and the comparison revealed a very good agreement between these two approaches even for very narrow pores provided both walls are in a partial
wetting/drying regime. The geometric interpretation of Kelvin's equation was also supported by constructing 2D density profiles. However, if the contact angle of at
least one wall is zero (or $\pi$), the predictions of the macroscopic Kelvin equation worsen significantly, especially as the pore width decreases. Nevertheless, a
substantial improvement of the Kelvin equation can be achieved by taking into account the effect due to the presence of wetting (or drying) films. The film thicknesses
that appear in this corrected Kelvin equation (\ref{kelvin_het2}) can be determined from equation~(\ref{ell}) provided the pore is wide enough ($L\gtrsim 50\,\sigma$, with
$\sigma$ being the molecular diameter).

\item We have then revisited a model of a capillary slit with competing walls. Here, the Kelvin equation as well as symmetry considerations dictate that any phase
coexistence must occur at $\mu=\mu_\text{sat}$. In this case, the liquid-vapour interface is parallel to the walls and can be pinned to either of the walls (in which case
the system is in a condensed or evaporated state) or can be unbounded from both walls around a halfway between the walls; the latter state is referred to as a soft mode
phase or a delocalized interface. Nature and the location  of a transition when  the interface unbinds (interface localized-delocalized transition) are determined by an
interplay between the wetting properties of the walls and the finite-size effects due to finite separation $L$ of the walls:

\begin{enumerate}

\item According to macroscopic arguments ($L\gg\sigma$), the transition occurs at $T^*=T_\text{w}=T_\text{d}$, assuming the walls are perfectly antisymmetric. If the walls exhibit continuous
wetting/drying transition at $T_\text{w}$, then $T^*$ is a critical temperature, below which either of the localized state is equally stable. As   $T\to T_\text{w}^-$, the liquid-gas
unbinds continuously from a given wall and moves towards the pore midpoint. Above $T_\text{w}$ only a single state corresponding to a delocalized interface is present. If the
walls undergo first-order wetting transition, then $T^*=T_\text{w}$ (and $\mu=\mu_\text{sat}$) is a triple point at which all three configurations coexist. These macroscopic
predictions were verified by DFT calculations for pores of widths $L\gtrsim50\,\sigma$.

\item As the pore width is reduced, $T^*$ does not coincide with $T_\text{w}$ anymore but becomes a function of $L$. The behaviour of $T^*(L)$ then strongly depends on the nature of
the wetting transition of the walls and can be described by mean-field mesoscopic arguments that capture the effect of the effective interaction between the liquid-gas
interface and the attractive tails of the wall potentials. For the walls exhibiting first-order wetting transition, the analysis shows that there are three temperature
regimes. For $T<T_\text{w}$, the only stable solutions correspond to a bounded interface for any values of $L$. For $T_\text{w}<T<T_\text{s}$, there is an interface localized-delocalized
transition for the wall separation $L^*(T)$ which decreases with an increasing $T$. Above $T_\text{s}$, there is no transition and there is a single configuration corresponding to
the delocalized interface. The spinodal temperature $T_\text{s}$ is characterised by a disappearance of the energy barrier in the binding potential and can be associated with the surface critical point of prewetting transition. The spinodal temperature determines the critical width $L_\text{c}=L^*(T_\text{s})$,
which is a minimal value of $L$ allowing  the interface localization-delocalization transition. Our DFT results suggest that $L_\text{c}$ is slightly below $10\,\sigma$, in a
reasonable agreement with the mesoscopic prediction as given by equation~(\ref{lc})  with the parameters $A$ and $B$ determined from equations~(\ref{A}) and (\ref{B}), which
yields $L_\text{c}\approx6\,\sigma$. Conversely, these results can also be interpreted such that for each $L_\text{c}<L<\infty$ there exists a temperature $T^*$ below which the fluid
undergoes capillary filling (as $\mu$ approaches $\mu_\text{sat}$ from below) or emptying (as $\mu$ approaches $\mu_\text{sat}$ from above) and above which no transition
is present and the interface is delocalized.

\item If the walls exhibit critical wetting at $T_\text{w}$ then, in contrast to the case of first-order wetting, the finite-size effects favour interface delocalization, so
that the interface localized-delocalized transition is shifted below $T_\text{w}$. For very large $L$, this shift is proportional to $L^{-1}$ with a (non-universal) temperature dependent amplitude that can be in principle determined
from a given molecular model and this result is consistent with the one obtained on the basis of finite-size scaling \cite{par_ev1}. For moderately large values of $L$ and the model with the binding potential
$W(\ell)=A(T)\ell^{-2}+B(T)\ell^{-3}+\cdots$, the mesoscopic analysis suggests that the phase behaviour can be characterized by a dimensionless parameter $\kappa=B/AL$, such that the localized-deloca\-lized transition occurs for
$\kappa=-2/3$ and $\kappa\approx-0.24$ with the delocalized interface being a stable solution within this interval.

\end{enumerate}

\item Within the microscopic DFT model that has been considered in this work, the walls exert long range (decaying as $z^{-3}$ for large $z$) potential while
the fluid-fluid interaction was taken to be only short-ranged. This particular choice of the interaction model requires some comments:

\begin{enumerate}

\item In terms of the study of interface localized-delocalized transition, this model has one advantage and one disadvantage. The great advantage is that the
competing walls can be made antisymmetric by simply choosing $\varepsilon_{\text{w}_1}=-\varepsilon_{\text{w}_2}$ and this antisymmetry is maintained regardless of the temperature.
This is because the only temperature dependent factor in both Hamaker constants (corresponding to the situation when the adsorbed film at the wall is a liquid or a gas
phase) is $(\rho_\text{l}-\rho_\text{g})$. Therefore, the Hamaker constants certainly vary with temperature but in  identical manner for both walls which allowed us to study the phase behaviour of a
fluid between antisymmetric walls for various temperatures. This is in contrast with the model used in reference~\cite{stewart}, where the fluid-fluid interaction is also
long-ranged which produces an extra temperature dependent factor in the respective Hamaker constants $A_\text{w1}\propto(\rho_\text{l}-\rho_\text{w})$ and $A_\text{w2}\propto(\rho_\text{g}-\rho_\text{w})$ that are
identical at only one particular temperature. On the other hand, a disadvantage of our model is that since the Hamaker constant is always positive, a possibility that the
walls undergo critical wetting is excluded, hence we could only test the cases when the walls exhibit first-order wetting.

\item Setting $\varepsilon_{\text{w}_1}=-\varepsilon_{\text{w}_2}$ makes the wall 2 purely repulsive, so that even though $A_1=A_2$ the walls cannot be viewed as perfectly
antisymmetric in the sense $T_\text{w}=T_\text{d}$ ($T_\text{d}$ is the drying temperature of the wall 2), because the wall 2 is completely dried at all temperatures for which vapour and liquid
may coexist. This broken (anti)symmetry is reflected in the binding potential for the capillary with competing walls which exhibits at most two minima, missing a local
minimum near wall 2. In contrast with the fully antisymmetric model (considered within the macro- and mesoscopic pictures), the temperature $T^*$ for first-order wetting
is not a triple temperature but a temperature appropriate to an ordinary first-order transition at which the liquid-gas interface jumps from the proximity of wall 1 to
the centre of the pore in some analogy to prewetting transition (in contrast to the latter, the transition takes place at $\mu=\mu_\text{sat}$ and the jump is determined
by $L$). Also note that below $T_\text{w}$, $\cos\theta_1+\cos\theta_2$ appearing in Kelvin's equation becomes negative, meaning that for $T<T_\text{w}$ a two phase coexistence
(capillary evaporation) occurs at $\mu>\mu_\text{sat}$. Finally, if the competing walls have different Hamaker constants, the interface localized-delocalized transition
is still possible but with the position of the delocalized interface near $\ell=L/[1+(A_2/A_1)^{1/3}]$ rather than~$\ell=L/2$.

\end{enumerate}
\end{enumerate}

\section*{Acknowledgements}
The financial support from the Czech Science Foundation, project number 13-09914S is acknowledged.

\ukrainianpart

\title{Фазові переходи плинів в гетерогенних порах}
\author{A. Малієвский\refaddr{label1,label2}}
\addresses{
\addr{label1} Відділ фізичної хімії, Празький університет хімічної технології, 166 28 Прага 6, Чеська Республіка
\addr{label2} Лабораторія аерозольної хімії і фізики, Інститут основ хімічних процесів, Академія наук, \\ 16502 Прага 6, Чеська Республіка
}

\makeukrtitle

\begin{abstract}
Ми вивчаємо фазову поведінку модельного плину, обмеженого двома  різними паралельними стінками у присутності далекосяжних (дисперсійних) сил. Передбачення, отримані на основі макроскопічних (геометричних) і мезоскопічних аргументів, порівнюються з числовими розв'язками  теорії функціоналу нелокальної густини. Розглянуто дві капілярні моделі.
У випадку капілярної моделі, що має дві (по-різному) адсорбуючі поверхні, прості геометричні аргументи приводять до узагальненого рівняння Кельвіна, яке дуже точно локалізує капілярну конденсацію, за умови, якщо обидві стінки лише частково змочуються. Якщо принаймні одна зі стінок знаходиться у режимі повного змочування, рівняння Кельвіна слід видозмінити за рахунок ефекту  товщини змочуючих плівок, включивши поправку Дєрягіна. У другій моделі розглянуто капіляр, утворений з двох конкуруючих стінок: одна з них схильна до змочування, а друга~--- не змочується.
У цьому випадку на границі розділу відбувається локалізований-делокалізований перехід при двофазному співіснуванні в об'ємі і при температурі $T^*(L)$, яка залежить від ширини пори  $L$. Аналіз, виконаний за допомогою теорії середнього поля, показує, що для стінок, які проявляють перехід змочування першого роду при температурі  $T_\text{w}$, $T_\text{s}>T^*(L)>T_\text{w}$, де температура спінодалі  $T_\text{s}$ може бути зв'язана з критичною температурою попереднього змочування, яка також визначає критичну ширину  пори, нижче якої локалізований-делокалізований перехід на границі розділу не відбувається.
Якщо ж стінки проявляють критичне змочування, тоді перехід зміщується  нижче за $T_\text{w}$, а  для моделі з потенціалом
$W(\ell)=A(T)\ell^{-2}+B(T)\ell^{-3}+\cdots$, де $\ell$~---  місцезнаходження границі розділу між рідиною і газом, перехід можна характеризувати за допомогою безрозмірного параметра $\kappa=B/(AL)$, в результаті чого конфігурація плину з делокалізованою границею розділу є стабільною в інтервалі між  $\kappa=-2/3$ і $\kappa\approx-0.23$.
\keywords капілярна конденсація, змочування, рівняння Кельвіна, адсорбція,  теорія функціоналу густини, теорія фундаментальної міри
\end{abstract}

\end{document}